
\def\lsim{\mathrel{\scriptstyle{\buildrel < \over \sim}}}
\magnification 1200
\baselineskip=17pt

\centerline{\bf MAGNETIC PROPERTIES OF THE SKYRMION GAS IN TWO
DIMENSIONS}
\vskip 50pt
\centerline{J. P. Rodriguez}
\smallskip
\centerline{\it Instituto de Ciencia de Materiales (CSIC),
Universidad Autonoma de Madrid,}

\centerline {\it Cantoblanco, 28049 Madrid, Spain.}
\centerline{\it and}
\centerline{{\it Dept. of Physics and Astronomy,
California State University,
Los Angeles, CA 90032.}\footnote*{Permanent address.}}

\vskip 30pt
\centerline  {\bf  Abstract}
\vskip 8pt\noindent
The classical ferromagnet is analyzed in two dimensions 
via a mean-field treatment of the
CP$^1$ model valid in the limit of low temperature  and  of
weak magnetic field.   At high skyrmion
densities, we find that the magnetization begins to decrease linearly with
temperature, but it then crosses over to a Curie-law at
higher temperature.  The comparison of these
results  with recent  Knight-shift experiments on the quantum
Hall state is fair. 
\bigskip
\noindent
PACS Indices: 73.40.Hm, 75.10.Hk, 75.40.Cx 
\vfill\eject

It is well known that the continuum ferromagnet supports topologically
non-trivial spin textures called skyrmions in two dimension.$^{1-3}$
In physical terms such skyrmions represent domain-wall loops within
which the magnetization is reversed.  The scale invariance
characteristic of the Heisenberg model in two dimensions indicates
that  the corresponding skyrmion solutions  have no preferred size, 
nor do they  interact, at zero temperature.  Quantum fluctuations
in the form of dynamical spin-wave excitations, however, break
this scale invariance, which means that
skyrmions interact  at non-zero temperature.  
Similar effects arise in the case
of the two-dimensional (2D) quantum antiferromagnet, but  at {\it zero}
temperature as well.$^{4}$

Recent theoretical and experimental work indicates that skyrmion
spin textures are tied to excess electronic charge with respect 
to a filled Landau
level  in the quantum Hall effect,$^{5-7}$ at which 
point the groundstate is a 2D  ferromagnet.  Motivated by
this as well as by general theoretical considerations, we shall study 
here the magnetic properties of a  gas of skyrmions
with a net topological charge density.  
Like it's electronic counterpart, the total skyrmion number
is necessarily quantized in integer values
due to the topological nature of the 2D
ferromagnet.$^{1-3}$  It is also
a conserved quantity,  which
justifies the restriction to constant skyrmion number
assumed throughout.  The magnetic
correlations, as well as the magnetization, are calculated as a function
of temperature and of external magnetic field via 
a meanfield analysis of the CP$^1$  model
for the {\it classical} ferromagnet.$^{2,3,8}$  Quantum
effects$^9$ related  to dynamical spin-wave excitations are therefore
neglected.   Notably, we 
obtain the following results in the limit of high skyrmion density:
(a) the magnetic correlation length
is set   by the inter-skyrmion separation; 
(b) the paramagnetic susceptibility displays a Curie law at {\it all}
temperatures due to   the disordering effect
of the skyrmion gas; and (c) the magnetization in fixed external field
crosses over from a
low-temperature linear decrease to the above Curie-law tail
at high temperature.  Although
the latter cross-over is weak (see Fig. 1), the agreement between the
theoretically predicted cross-over temperature 
with that observed in recent Knight-shift
measurements$^7$ on the quantum Hall effect is surprisingly good.

{\it CP$^1$ Model.} We now consider the thermodynamics of the
classical  2D ferromagnet in the presence of a net
skyrmion density.
The energy functional of the corresponding continuum 
Heisenberg model may be written in
the CP$^1$ form as
$$ E = 2 s^2 J\int d^2r |(\vec\nabla - i\vec a) z|^2, \eqno (1)$$
where $z = (z_-, z_+)$ is a complex valued doublet field constrained
to unit modulus,
$$\bar z z = 1, \eqno (2)$$
and where the vector potential $\vec a$ is tied to the $z$ fields
by
$$ \vec a = i(\vec\nabla\bar z) z.\eqno (3)$$
Here $s$ denotes the microscopic spin.
The $z$ fields are related to the normalized magnetization,
$\vec m$,  of the Heisenberg model by
$$\vec m = \bar z \vec\sigma z,\eqno (4)$$
where $\vec\sigma$ denotes the Pauli matrices,
while the spin stiffness  (1) is given by $\rho_s = s^2 J$.
Using Eqs. (2)-(4), it can be shown that the constrained
($|\vec m|^2 =1$) energy functional $E = {1\over 2}\rho_s
\int d^2 r |\vec\nabla\vec m|^2$ is equal to  
that of the CP$^1$ model  (1).$^{2,3,8}$  Similarly, it can be shown
that the skyrmion density is equal to the density of fictitious
magnetic flux, $(\partial_x a_y - \partial_y a_x)/2\pi$.

To proceed further, we now write the Gibbs distribution,
$Z = \int {\cal D} \vec m\, \delta(|m|^2 - 1) e^{-E/k_B T}$,
in the presence of external magnetic field $H$ along the
$z$-axis
as
$$ Z = \int {\cal D}z {\cal D}\bar z {\cal D}\vec a {\cal D}\lambda\,
{\rm exp} \Biggl\{-2\beta\int d^2 r[|(\vec\nabla - i\vec a) z|^2
+i\lambda(|z|^2 - 1)
+ b_Z(|z_-|^2 - |z_+|^2)]\Biggr\}, \eqno(5)$$
where we define $\beta = s^2 J/k_B T$
and $b_Z = n E_Z/4 s^2 J$, with spin density $n$ and
Zeeman energy splitting per spin $E_Z = g\mu_B H$.  Here
integration over the Langrange multiplier field $\lambda (\vec r)$
enforces    constraint (2), while integration over the now unconstrained
vector potential field $\vec a$
in the Coulomb gauge, $\vec\nabla\cdot\vec a = 0$,
recovers constraint (3).
We now make the basic  approximations  of the paper,
which are (i) to enforce constraint (2) only on average over the entire
2D plane, and (ii) to neglect spatial fluctuations 
in the skyrmion density;$^3$ i.e., the $z$ fields above are integrated
out first  with the presumption of  a homogeneous Langrange multiplier field
$\lambda(\vec r) = \lambda_0$  and of a homogeneous fictitious
magnetic field $b_S = (\partial_x a_y - \partial_y a_x)$.
The latter is of course  equal to the net skyrmion density multiplied by 
a factor of $2\pi$.  The saddle-point condition 
$ {\partial\over{\partial\lambda_0}}{\rm ln}\, Z = 0 $ is equivalent to
the meanfield constraint 
$1 = \langle\bar z_- z_-\rangle + \langle\bar z_+ z_+\rangle$, where
$$\langle\bar z_{\pm} z_{\pm}\rangle 
=  (2\beta)^{-1} V^{-1} {\rm tr}[-(\vec\nabla - i\vec a)^2 +i\lambda_0
\mp  b_Z]^{-1} \eqno (6)$$ 
are the respective averages.  But since the
spectrum of the Hermitian operator $- (\vec\nabla - i\vec a)^{2}$
is just that of Landau levels with energies $2b_S(n+{1\over 2})$,
each with a degeneracy per area $V$ of $b_S/2\pi$, we obtain the
meanfield equation
$$2\pi\beta = {1\over 2}\sum_{n=0}^{\infty}{b_S\over{2b_S n + \xi^{-2} +
b_Z}} + {1\over 2}\sum_{n=0}^{\infty}{b_S\over{2b_S n + \xi^{-2} -
b_Z}}\eqno (7)$$
for the correlation length $\xi$ set by
$\xi^{-2} = i\lambda_0 + b_S$.  Similarly, we obtain the meanfield
expression
$$m_z = (2\pi\beta)^{-1}\Biggl( 
{1\over 2}\sum_{n=0}^{\infty}{b_S\over{2b_S n + \xi^{-2} -
b_Z}} - {1\over 2}\sum_{n=0}^{\infty}{b_S\over{2b_S n + \xi^{-2} +
b_Z}}\Biggr)\eqno (8)$$
for the normalized magnetization 
$m_z = \langle\bar z_+ z_+\rangle - \langle\bar z_- z_-\rangle$ 
along the $z$ direction.

Consider now the ferromagnetic regime, $b_S\rightarrow 0$, where
the concentration of skyrmions is dilute with respect to the
magnetic correlation length, but is not zero.  
The sums in Eqs. (7) and (8) may then
be converted into integrals,  resulting in
$2\pi\beta =  {\rm ln}\, [k_0 (\xi^{-4} - b_Z^2)^{-1/4}]$ and
$m_z = (2\pi\beta)^{-1} 
{\rm ln}\, [(\xi^{-2} + b_Z)/(\xi^{-2} - b_Z)]^{1/4}$ respectively.  
Here $k_0$ represents the momentum cut-off corresponding to  the former 
integral. We therefore obtain the familiar result
$\xi = (\xi_0^{-4} + b_Z^2)^{-1/4}$ for the correlation length,$^3$
where the zero-field ferromagnetic correlation length
$\xi_0 = k_0^{-1} e^{2\pi\beta}$ diverges exponentially at
low temperature.   The
magnetization is paramagnetic in the limit of weak external
field, $b_Z\rightarrow 0$, following $m_z = \chi_Z b_Z$ 
with $\chi_Z = (4\pi\beta)^{-1}\xi_0^2$.  
In general, it is given by the expression
$m_z = 1 - (2\pi\beta)^{-1}{\rm ln}[k_0(\xi^{-2} + b_Z)^{-1/2}]$,
which means that the normalized magnetization begins to saturate
logarithmically as external field increases.
The continuum limit then breaks down at
$k_0 b_Z^{-1/2}\lsim 1$, where $m_z$ 
is pathological and exceeds unity.$^{10}$
The present results correspond to those of the quantum ferromagnet
in the renormalized classical  regime.$^9$  
The former high-field catastrophe is avoided in that case,
however, where
the system evolves into a quantum activated regime instead.
Also, it can be easily shown that 
$\langle z_{\pm} (\vec r) \bar z_{\pm} (\vec r\,^{\prime})\rangle
\sim {\rm exp}[-|\vec r -\vec r\,^{\prime}|(\xi^{-2}\mp b_Z)^{1/2}]$
at long distance $|\vec r - \vec r\,^{\prime}|$, which by
Eq. (4) indicates that $(\xi^{-2} - b_Z)^{-1/2}$ gives
the magnetic correlation length.
Last, if $\delta b_S (\vec r)/2\pi$ denotes local fluctuations in the skyrmion
density, while  $\delta\lambda (\vec r)$ denotes the corresponding
fluctuations of the Langrange multiplier field,
then it can be shown that the Gibbs distribution associated
with such fluctuations is given by$^{3,8}$
$$Z_{2} = 
\int {\cal D}  b_S {\cal D} \lambda\, {\rm exp}
\Biggl\{-{1\over 2}\int d^2r\, [\chi (\delta b_S)^2
+2i\sigma(\delta b_S)(\delta\lambda)
+\epsilon(\vec\nabla\delta\lambda)^2]\Biggr\}\eqno (9)$$
in the long-wavelength limit (relative to the inter-skyrmion separation),
where  $\chi\sim\xi_0^2$, $\sigma = 2 \beta/b_S $ and 
$\epsilon = \beta/b_S^2$ (see the Appendix).
Hence, the direction of steepest descent for gradients of the Lagrange
multiplier field is along the {\it real}
axis in the presence of a net skyrmion density.
Notice also that homogeneous
fluctuations ($\vec\nabla\delta\lambda = 0$)
of the Langrange multiplier  field are in general only marginally stable
in such case.   Fluctuations, 
$V^{-1}\langle (\int d^2r\, \delta b_S/2\pi)^2\rangle \sim  \xi_0^{-2}$,
in the total skyrmion number
are exponentially
suppressed at low temperature, on the other hand.
The latter thermally
activated temperature
dependence is then of course
entirely consistent with the fact that $E_S = 4\pi s^2J$ is the energy
cost of a single skyrmion.$^{1}$  

Consider next the skyrmion-rich limit, $b_S\rightarrow\infty$, where
the separation between  neighboring skyrmions is much less than the 
ferromagnetic correlation length, $\xi_0$.
In that case the $n=0$ terms corresponding to the lowest
Landau level dominate the sums in Eqs. (7) and (8), which results
in the meanfield equations
$2\pi\beta = b_S\xi^2/(1-b_Z^2\xi^4)$ and $m_z = \xi^2 b_Z$
respectively.  Solving the former relationship explicitly,
we obtain the simple expression
$$m_z = (1+\bar t^2)^{1/2} - \bar t \eqno (10)$$
for the normalized magnetization, where 
$$\bar t = {k_B T\over {E_Z}} {b_S\over{\pi n}} \eqno (11)$$
is the reduced temperature.  The magnetization therefore decreases
linearly from unity at low temperature
and/or high magnetic field, while it follows a
Curie law, $m_z\cong {1\over 2} \bar t^{-1}$, at high temperature
and/or low magnetic field.
(It is interesting to note that a similar $H/T$ dependence is
expected  for the magnetization of the quantum ferromagnet 
in the high-temperature quantum-critical regime, yet     
in the absence of a net skyrmion density.$^{9}$)
The effective area $A_S(T)$ of a single 
skyrmion can be obtained from the comparison of 
Eq. (10) at low temperature  with 
the naive formula
$m_z = 1 - 2 A_S (b_S/2\pi)$
for the magnetization obtained by counting the number of reversed spins,
$S = n A_S$, per skyrmion.
This yields an effective
area of $A_S = (k_B T/E_Z) n^{-1}$ that results from
the opposition of the Zeeman energy, which
tends to decrease  the number of
reversed spins, with the entropic part of the free energy,
which tends to increase the number of
reversed spins.
The cross-over between these two regions naturally  occurs  at 
$\bar t = 1$ (see Fig. 1), where the effective size of the skyrmion
is by definition on the order of the inter-skyrmion separation.  
Also, the correlation function for the
$z$ fields has the gaussian form
$|\langle z(\vec r) \bar z(\vec r\,^{\prime})\rangle|^2
= {\rm exp}(-{1\over 2}|\vec r - \vec r\,^{\prime}|^2 b_S)$ in above
lowest-Landau-level approximation, which means that the magnetic
correlation length is limited by the inter-skyrmion separation
for all values of external magnetic field, $b_Z$, 
in this regime. 
This magnetic correlation length therefore also limits the  size
of a single skyrmion to the inter-skyrmion
separation.  Last, the long-wavelength 
fluctuation corrections to the present
saddle-point (9) are identical to that of the skyrmion-poor
regime, with the exception that
$\chi = 2 \beta/b_S = \sigma$ (see the Appendix).
This means
that fluctuations, $V^{-1}\langle (\int d^2r\, \delta b_S/2\pi)^2\rangle
\sim  \chi^{-1}$, in the total skyrmion number of the system are
much larger in this regime.  

{\it Quantum Hall Effect.}  Let us apply the above results to the
quantum Hall effect in the vicinity of unit filling,
where it has been suggested that the skyrmion density 
$b_S/2\pi$ is
equal to the excess charge density $\delta n$;$^{5-7}$ i.e., the electronic
filling factor,
$\nu = 2\pi  l_0^2 n$, is related to the skyrmion density
by $|\nu - 1| =  l_0^2 b_S$, where $l_0$ denotes the
magnetic length in fixed magnetic field.  Using these relations,
we then obtain the simple expression
$$\bar t = 2 {|\nu - 1|\over{\nu}} {k_B T\over{E_Z}}\eqno (12)$$
for the reduced temperature (11).  This immediately defines a
cross-over temperature between  the low-temperature regime
with linearly decreasing magnetization and the high-temperature
Curie-law regime (see Fig. 1) given by
$k_B T_* = {1\over 2} |1 - \nu^{-1}|^{-1} E_Z$,
which is notably independent of the spin stiffness.
Recent Knight shift measurements exhibit a temperature dependence 
that is qualitatively
similar to that predicted by Eq. (10), yet with a more pronounced
cross-over.$^7$  Such  measurements were  conducted in magnetic fields
of $H\cong 7\,{\rm T}$ 
with a Zeeman energy of $E_Z = 0.2\,{\rm meV}$,  
and at filling fractions of
$\nu = 0.88$ and $\nu = 1.2$.
The present theory for  
the skyrmion rich regime then predicts cross-over temperatures of
$T_* = 8\,{\rm K}$ and $T_* = 7\,{\rm K}$, respectively.
The sharp cross-overs observed experimentally  at roughly $9\,{\rm K}$
in both cases compare quite      well with these estimates.
But do  such Knight-shift measurements actually lie within the
skyrmion-rich limit discussed here?  
Inspection of Eq. (8) for the magnetization
indicates that the skyrmion density
must satisfy $b_S > {\rm max}(b_Z, \xi_0^{-2})$ for
this to be the case.  Presuming that
the momentum cut-off is set by the magnetic length,$^{11}$
$k_0\sim l_0^{-1}$, then the latter is equivalent to satisfying
conditions (a) $b_S > b_Z$, and (b) $|1-\nu| > e^{-E_S/k_B T}$,
where $E_S = 4\pi \rho_s$ is the energy cost of creating a skyrmion
in the ferromagnet.  Standard estimates for 
the spin-stiffness $\rho_s$ of the quantum Hall state
at unit filling$^{5,9}$ place this energy at
$E_S/k_B\sim 40\,{\rm K}$ 
in a $7\,{\rm T}$ field, which means that condition (b) is
easily satisfied at  temperatures much less than this
scale.  Also, 
employing Eq. (11) at the cross-over ($\bar t = 1$)
implies that condition (a) is satisfied as well,
since $T_*\lsim 10\,{\rm K}$
falls substantially below the previous scale.
Last, the present classical treatment of the ferromagnet 
is generally valid in the renormalized classical regime,$^9$
$E_Z < k_B T < E_S$, which spans the majority of the experimental
temperature range.$^7$

In conclusion, although the 
magnetization (10) obtained here 
for the classical 2D ferromagnet in the presence of
a net skyrmion concentration compares favorably
with recent Knight shift experiments on the quantum Hall state,
it   consistently overestimates the latter.$^7$
From the theoretical side, a number of effects could be responsible
for this discrepancy.  First, it has been noted
that fluctuations with respect to the present meanfield 
treatment of the classical ferromagnet are relatively large
in the presence of skyrmions.  
More generally, it is also known that
fluctuations in the skyrmion density 
act to confine the $z$ and $\bar z$ fields
in strictly two dimensions.$^{8}$  
Such effects have not been accounted for
here.  Second, 
the inclusion of quantum mechanical spin-wave excitations 
will further decrease the magnetization.  Last, both quantum
fluctuations$^4$ as well as the direct Coulomb interaction$^{12}$
tend to crystallize the skyrmion gas.  
Accounting for  
all of the above-mentioned effects remains as a problem for the
future.

The author has benefited from numerous 
discussions with L. Brey and A. Somoza.
This work  was supported in
part by  the Spanish Ministry for Science and Education, as well
as by National
Science Foundation grant DMR-9322427.

\vfill\eject

\centerline {\bf Appendix}
\vskip 16 pt
We shall now compute the fluctuation corrections (9) to the 
present meanfield
approximation.  If we take the Landau gauge,
$\vec a = (0, b_S x)$, then the eigenstates of the
Hermitian operator $-(\vec\nabla - i\vec a)^2 +i\lambda_0$ have the
usual form $\langle \vec r|n,q\rangle = \langle x^{\prime}|n\rangle e^{-iqy}$,
with eigenvalues $\varepsilon_n = 2 b_S n + \xi^{-2}$.  Here
$|n\rangle$ denotes the eigenstates of the corresponding Harmonic
oscillator while $x^{\prime} = x -  b_S^{-1} q$.  If we now
identify the variation in the vector potential by
$\delta b_S = \partial_x\delta a_y - \partial_y\delta a_x$, then 
fluctuations correct the Gibbs distribution by a factor of
$$\eqalignno{
{\rm exp}\Biggl( 
-{1\over 2}V^{-1}\sum_{\vec k}\Bigl\{  
[\Pi_{ij}^+(\vec k) + \Pi_{ij}^-(\vec k)]
\delta a_i(\vec k) & \delta a_j(-\vec k) 
+ 2i[\Pi_{0j}^+(\vec k) + \Pi_{0j}^-(\vec k)]
\delta a_j(\vec k)  \delta \lambda (-\vec k) +\cr
&
+[\Pi_{00}^+(\vec k) + \Pi_{00}^-(\vec k)]
\delta\lambda(\vec k)\delta\lambda(-\vec
k)\Bigr\}\Biggr), & (A1)\cr
}$$
where $\Pi_{ij}^{\pm} (\vec k)$,  $\Pi_{00}^{\pm}(\vec k)$ 
and $\Pi_{0j}^{\pm}$ are respectively
the high-temperature limits of the Kubo formula, the Lindhard
function and the hybrid Kubo-Linhard function
corresponding to the $z_{\pm}$ field.  
Also, $\delta a_i(\vec k)$ and $\delta\lambda(\vec k)$ denote the
appropriate Fourier components.
One may presume that
$\vec k$ be parallel to the $x$-axis without any loss of
of generality, in which case the polarizabilities are given explicitly by
$$\eqalignno{
\Pi_{ij}^{\pm}(\vec k) = & V^{-1}
\sum_{n,q}{2 \delta_{ij}\over{\varepsilon_n\mp b_Z}}
-
V^{-1}
\sum_{n\neq n^{\prime},q} {\langle n,q|v_i(\vec k)|n^{\prime},q\rangle\over
{\varepsilon_n\mp b_Z}} {\langle n^{\prime},q|v_j(-\vec k)|n,q\rangle\over
{\varepsilon_{n^{\prime}}\mp b_Z}}, & (A2)\cr
\Pi_{0j}^{\pm}(\vec k) = & 
V^{-1}
\sum_{n\neq n^{\prime},q} {\langle n,q|v_j(\vec
k)|n^{\prime},q\rangle\over
{\varepsilon_n\mp b_Z}} {\langle n^{\prime},q|e^{-ikx}|n,q\rangle\over
{\varepsilon_{n^{\prime}}\mp b_Z}}, & (A3)\cr
\Pi_{00}^{\pm}(\vec k) = &  V^{-1}
\sum_{n\neq n^{\prime},q} {\langle n,q|e^{ikx}|n^{\prime},q\rangle\over
{\varepsilon_n\mp b_Z}} {\langle n^{\prime},q|e^{-ikx}|n,q\rangle\over
{\varepsilon_{n^{\prime}}\mp b_Z}}, & (A4)\cr
}$$
where $\vec v (\vec k)
= 2  e^{i{1\over 2}kx}(i\vec\nabla + \vec a)
e^{i{1\over 2}kx}$
is the modified velocity operator.  In general, these polarizabilities
have the form 
$\Pi_{00}^{\pm}(\vec k) = \epsilon_{\pm} k^2$,
$\Pi_{0y}^{\pm}(\vec k) = \sigma_{\pm} ik$ and
$\Pi_{yy}^{\pm}(\vec k) = \chi_{\pm}k^2$ in the long-wavelength
limit, while the remaining ones are identically
zero due to gauge invariance.  
A direct application of (A3) and (A4) 
then yields
$\sigma_{\pm} = b_S^{-1}  V^{-1}
\sum_{n,q} (\varepsilon_n\mp b_Z)^{-1}$
and $\epsilon_{\pm}=b_S^{-2} {1\over 2} V^{-1}
\sum_{n,q} (\varepsilon_n\mp b_Z)^{-1}$.
Employing mean-field Eq. (7),
we then obtain that the net fluctuation susceptibilities
(9) are given by $\sigma = \sigma_+ + \sigma_- =  2 \beta/b_S$ and
$\epsilon = \epsilon_+ + \epsilon_- =     \beta/b_S^2$
in general.  Similar calculations for (A2) also demonstrate 
ultimately that
$\chi = 2 \beta/b_S = \sigma$ in the skyrmion-rich limit,
$b_S\rightarrow \infty$.  This identity
is  suggested by the present mean-field analysis,
which depends crucially on the combination
$\xi^{-2} = i\lambda_0 +b_S$ [see Eq. (9)].

\vfill\eject

\centerline{\bf References}
\vskip 16 pt
\item {1.} A.A. Belavin and  A.M. Polyakov, Pis'ma
Zh. Eksp. Teor. Fiz. {\bf 22}, 503 (1975) [JETP Lett. {\bf 22}, 245
(1975)].

\item {2.} R. Rajaraman, {\it Solitons and Instantons}
(North-Holland, Amsterdam, 1987).

\item {3.} A.M. Polyakov, 
{\it Gauge Fields and Strings} (Harwood, New York, 1987).

\item {4.} J.P. Rodriguez, Phys. Rev. B {\bf 39}, 2906 (1989);
{\it ibid.} {\bf 41}, 7326 (1990).

\item {5.} S.L. Sondhi, A. Karlhede, S.A. Kivelson, and E.H. Rezayi,
Phys, Rev. B {\bf 47}, 16419 (1993).

\item {6.} H.A. Fertig, L. Brey, R. C\^ ot\' e, and
A.H. MacDonald, Phys. Rev. B {\bf 50}, 11018 (1994).

\item {7.} S.E. Barret, G. Dabbagh, L.N. Pfeiffer, K.W. West, and R.
Tycko, Phys. Rev. Lett. {\bf 74}, 5112 (1995).

\item {8.} E. Witten, Nucl. Phys. {\bf B149}, 285 (1979).
 
\item {9.} N. Read and S. Sachdev, Phys. Rev. Lett. {\bf 75},
3509  (1995).

\item {10.} Similar behavior for the magnetization is displayed
by the 2D spherical model [see for example E. Br\' ezin and J. Zinn-Justin,
Phys. Rev. B {\bf 14}, 3110 (1976), Eq. (44)].

\item {11.} A.G. Green, I.I. Kogan and A.M. Tsvelik, Phys. Rev. B
(1996).

\item {12.} L. Brey, H.A. Fertig, R. C\^ ot\' e and A.H. MacDonald,
Phys. Rev. Lett. {\bf 75}, 2562 (1995).

\vfill\eject
\centerline{\bf Figure Caption}
\vskip 20pt
\item {Fig. 1.}   Shown is mean-field result (10) for the
magnetization vs. temperature of the classical 2D ferromagnet in the limit
of high skyrmion density.  The pronounced Curie-law tail 
notably weakens the  cross-over at $T_*$, which should lie well
below $4\pi\rho_s/k_B$.

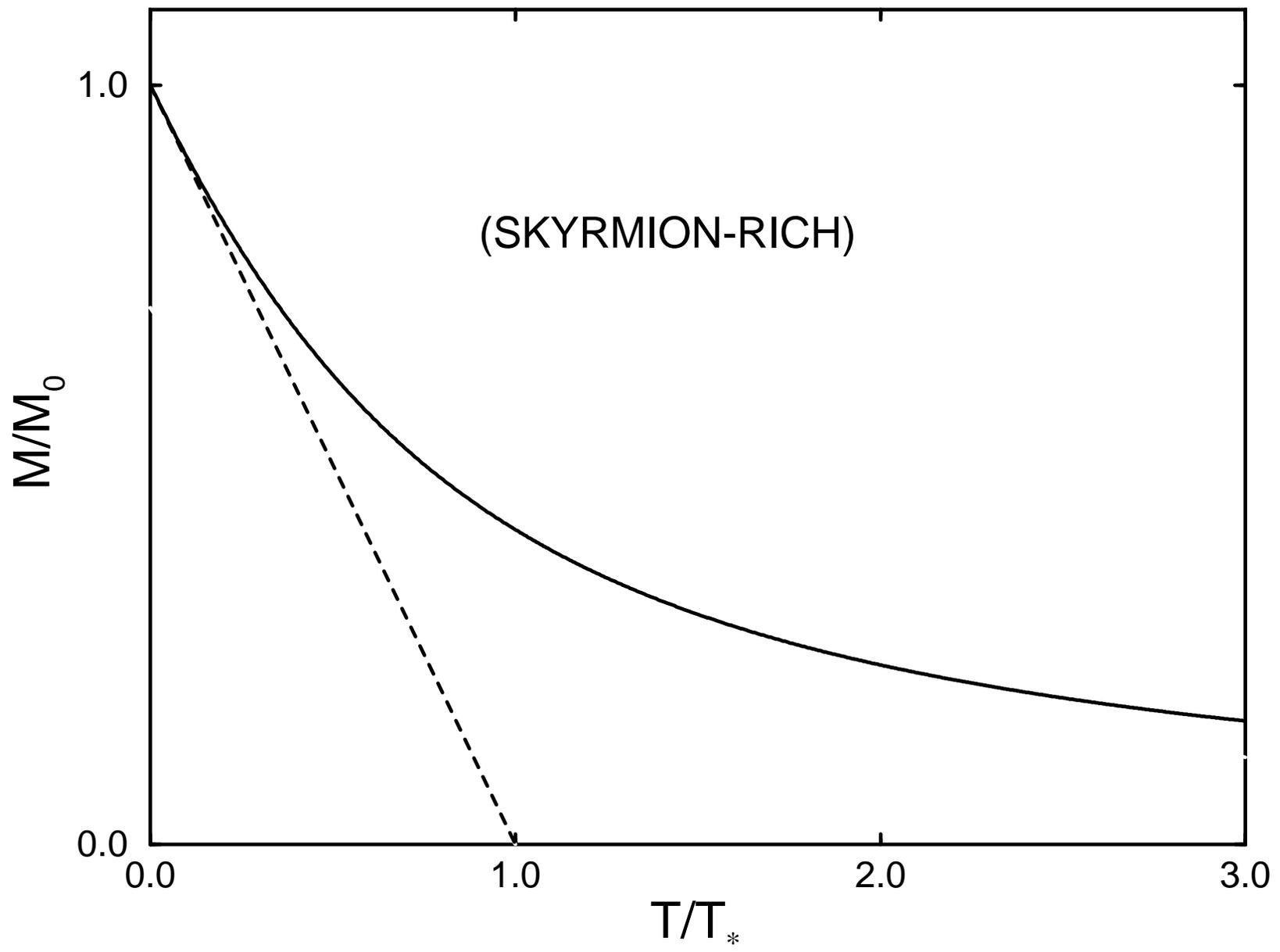
\end